\documentclass{IEEEtran}
\usepackage{graphicx}
\usepackage{amsmath,amsthm,amssymb}
\usepackage{stmaryrd}
\usepackage{booktabs}
\usepackage{hyperref}
\usepackage{multirow}
\usepackage{color}
\usepackage{enumitem}
\usepackage{cite}
\usepackage[ruled,linesnumbered]{algorithm2e}
\usepackage{siunitx}
\begin{document}

\title{Surrogate Modeling with Low-Rank Function Representation for Electromagnetic Simulation}

\author{Mingze Sun, Liang Li, Xile Zhao, Zheng Tan, Yulu Hu, Xing Li, Bin Li%
\thanks{M. Sun is with the Yingcai Honors College, University of Electronic Science and Technology of China (UESTC). L. Li and X. Zhao are with the School of Mathematical Sciences, UESTC. Z. Tan, Y. Hu, X. Li and B. Li are with the School of Electronic Science and Engineering, UESTC. The corresponding author is Liang Li (plum.liliang@gmail.com, plum\_liliang@uestc.edu.cn).}%
\thanks{Manuscript received February 10, 2026.}}

\markboth{IEEE TRANSACTIONS ON MICROWAVE THEORY AND TECHNIQUES,~VOL.~X, NO.~X, XXX~2026}%
{Sun \MakeLowercase{\textit{et al.}}: LOW-RANK FUNCTION REPRESENTATION FOR ELECTROMAGNETICS}

\maketitle

\begin{abstract}
High-fidelity electromagnetic (EM) simulations are indispensable for the design of microwave and wave devices, yet repeated full-wave evaluations over high-dimensional design spaces are often computationally prohibitive. While neural surrogates can amortize this cost, learning high-dimensional EM response mappings remains difficult under limited simulation budgets due to strong and heterogeneous parameter couplings. In this work, we introduce low-rank tensor function representations as a principled surrogate modeling paradigm for EM problems and provide a systematic study of representative low-rank formats, including Tucker-style low-rank tensor function representation (LRTFR) as well as neural functional tensor-train (TT) and tensor-ring (TR) baselines. Building on these insights, we propose a pairwise low-rank tensor network (PLRNet) that uses learnable pairwise interaction factors over compact coordinate-wise embeddings.
Experiments on representative EM surrogate tasks demonstrate that the proposed framework achieves a more favorable overall trade-off between accuracy, robustness, and parameter efficiency, with stable optimization in high-dimensional regimes.
\end{abstract}


\section{Introduction}
\IEEEPARstart{E}{lectromagnetic} simulation is a cornerstone for the design and analysis of modern microwave, antenna, and electromagnetic wave devices\cite{bila2000full,pietrenko2024low,kalantari2024efficient}. By numerically solving Maxwell’s equations, classical computational electromagnetics (CEM) methods, such as the finite-difference time-domain (FDTD) method\cite{yee1966numerical}, the finite element method (FEM)\cite{mur2003finite,li2013solution}, and the method of moments (MoM)\cite{song1997multilevel,li2013solution}, enable accurate prediction of scattering, radiation, and field distributions without the cost and inflexibility of repeated physical experiments. However, for large-scale, high-frequency, or strongly parameterized configurations, these solvers can be computationally prohibitive\cite{koziel2013physics,song1997multilevel}. Each new design typically requires geometry processing, meshing, and solving large linear/nonlinear systems, and downstream workflows such as parametric sweeps, inverse design, yield analysis, and uncertainty quantification may demand thousands to millions of repeated simulations\cite{adamczyk2014high}. This computational burden has become a major bottleneck for rapid EM prototyping and data-driven EM design.

Neural-network-based surrogate models offer a promising route to amortize this cost by learning a direct mapping from design parameters to quantities of interest\cite{grossmann2024can,meng2020ppinn,ohira2022surrogate,khan2024deep,zhang2021maxwell}. Typical inputs include geometric variables, material properties, boundary/excitation conditions, and frequency; outputs can be S-parameters, radar cross section (RCS), return loss, dispersion characteristics, and field patterns\cite{peurifoy2018nanophotonic}. Once trained, a surrogate can provide millisecond-level inference, enabling interactive prediction and accelerating gradient-free optimization loops\cite{lim2022maxwellnet,malkiel2018plasmonic,li2020fourier,lu2021learning}. Despite this progress, building accurate and scalable surrogates remains challenging in realistic EM settings. EM responses often depend on many coupled parameters, and the effective dimensionality of the design space can be high. In such regimes, simply increasing the width and depth of a generic multilayer perceptron (MLP), or adopting off-the-shelf architectures, frequently leads to over-parameterization, unstable training, and degraded generalization under limited simulation budgets, which reflects the curse of dimensionality. The central question is therefore how to design surrogate architectures that are simultaneously expressive, parameter-efficient, and robust for high-dimensional EM mappings.

A physically and empirically grounded clue is that many high-dimensional parametric EM responses exhibit substantial redundancy.
Smooth dependence on individual variables, implicit physical constraints, and correlated geometric effects often imply that the response manifold has a compact structure.
As a result, the mapping can often be well approximated by low-rank representations \cite{abbas2011characterization,zhang2016big}.
When responses are sampled on a Cartesian product grid, the evaluations form a high-order tensor, and many practical tensors lie near low-rank tensor families \cite{kolda2009tensor}.
This observation has motivated a broad class of low-rank tensor decompositions, including CP, Tucker, tensor-train (TT), and tensor-ring (TR), for compression and representation learning \cite{kolda2009tensor,oseledets2011tensor}.

Recently, Luo \emph{et al.}\ introduced the Low-Rank Tensor Function Representation (LRTFR) framework \cite{luo2023low}.
LRTFR provides a concrete way to carry the low-rank tensor viewpoint to continuous surrogate learning.
It maps each input coordinate to a learnable latent factor, and then couples these factors using a Tucker-type low-rank core.
This factorized inductive bias is appealing for EM surrogate modeling, because it can exploit redundancy when simulation data are expensive and limited.
We therefore adopt LRTFR as a natural starting baseline for EM surrogates.

However, Tucker-type coupling can become a bottleneck as the input dimension grows.
The parameter and memory cost of the Tucker core grows exponentially with the number of dimensions, which makes it difficult to maintain expressive ranks in high-dimensional design spaces \cite{cichocki2015tensor}.
Moreover, EM responses often exhibit non-uniform interactions across variables.
Some parameter groups are strongly coupled, while others are weakly coupled or behave differently across operating regimes.
A single global core couples all modes through the same bottleneck, and it may represent such heterogeneous interactions inefficiently \cite{khrulkov2017expressive}.

These observations motivate a broader view of low-rank tensor function representations.
The central idea is to model a multivariate function using coordinate-wise latent factors, and to couple these factors through a structured low-rank form.
From this perspective, Tucker coupling is one option, and tensor-network couplings such as TT and TR provide scalable alternatives in higher dimensions \cite{oseledets2011tensor,oseledets2010tt}.
In this work, we therefore consider multiple representative low-rank formats as structured baselines, including Tucker-type LRTFR and TT/TR variants, to clarify how different coupling structures behave on EM surrogate tasks.

Building on these insights, we identify two design motivations.
First, since low-rank is a family of structured representations, it is important to systematically compare representative low-rank tensor function formats for EM surrogate learning, beyond Tucker-type coupling alone.
Second, an EM-oriented surrogate should be scalable while modeling cross-parameter interactions in a flexible way.
To this end, we propose PLRNet, a Pairwise Low-Rank Tensor Network that models cross-parameter dependence through learnable pairwise interaction factors built on compact coordinate-wise embeddings.
By distributing interaction modeling across parameter pairs, PLRNet avoids a single global coupling bottleneck while retaining a low-rank structure that scales favorably with input dimension.

We evaluate PLRNet against a standard MLP surrogate \cite{rumelhart1986learning} and representative low-rank baselines, including LRTFR \cite{luo2023low} and TT/TR variants, on several EM surrogate benchmarks.
Across these tasks, PLRNet achieves improved prediction accuracy and parameter efficiency, with stable optimization in high-dimensional regimes.

In summary, this work makes the following contributions:
\begin{enumerate}[label=\textbf{\roman*})]
    \item \textbf{Paradigm introduction and baselines:} We introduce low-rank tensor function representations into electromagnetic surrogate modeling and implement representative formats, including Tucker-based LRTFR and tensor-network couplings based on tensor-train and tensor-ring, as structured baselines. This provides a methodological reference for high-dimensional EM surrogate learning.
    \item \textbf{EM-tailored architecture:} We propose PLRNet, a pairwise low-rank tensor network that captures diverse parameter interactions via learnable pairwise factors. PLRNet achieves improved accuracy and parameter efficiency, with robust optimization on EM benchmarks.

\end{enumerate}

\section{Low-Rank Tensor Function Representations as Baselines}
\label{sec:lowrank_baselines}

A major difficulty in high-dimensional electromagnetic (EM) surrogate modeling is to balance expressiveness and scalability under limited simulation budgets.
While generic MLPs can approximate complex mappings, their parameter count and optimization behavior often deteriorate as the number of input dimensions increases.
This motivates surrogate architectures with explicit low-rank structural priors.
In this work, we treat low-rank tensor function representations as a family of structured surrogates, distinguished mainly by how they couple different input coordinates.
We therefore introduce and implement several representative low-rank formats as baselines, including Tucker-type coupling (LRTFR) and tensor-network couplings (TT and TR), which will later motivate the design of PLRNet.

\subsection{Tucker-type coupling: LRTFR}
\label{subsec:lrtfr}

LRTFR~\cite{luo2023low} starts from a classical observation: when a multivariate function is sampled on a Cartesian product grid, its values form a high-order tensor, and many practical tensors admit accurate low-rank approximations.
To extend this low-rank perspective to continuous function modeling, LRTFR combines coordinate-wise latent factors with a Tucker-type coupling mechanism.

Specifically, for a scalar response $y = F(\boldsymbol{x})$ with $\boldsymbol{x}=(x_1,\dots,x_N)$, LRTFR models the surrogate as
\begin{equation}
  \hat F(\boldsymbol{x})
  =
  \left\langle
  \mathcal{C},\,
  f_{\theta_1}(x_1)\otimes f_{\theta_2}(x_2)\otimes\cdots\otimes f_{\theta_N}(x_N)
  \right\rangle,
  \label{eq:lrtfr_tucker}
\end{equation}
where $\mathcal{C}\in\mathbb{R}^{r_1\times\cdots\times r_N}$ is a learnable core tensor,
$f_{\theta_i}:\mathbb{R}\rightarrow\mathbb{R}^{r_i}$ maps the $i$-th scalar input to an $r_i$-dimensional latent vector,
and $\langle\cdot,\cdot\rangle$ denotes the Frobenius inner product.

\begin{figure}[!t]
  \centering
  \includegraphics[width=0.95\linewidth]{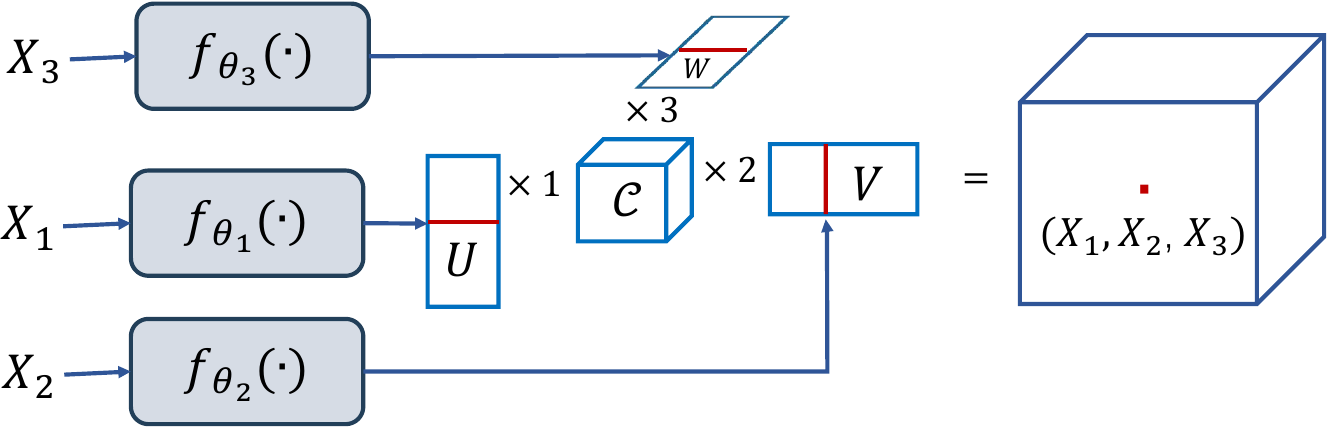}
  \caption{Tucker-type low-rank coupling in LRTFR (3D example): each scalar input is mapped to a latent vector, and the output is obtained by contracting these vectors with a global core tensor $\mathcal{C}$.}
  \label{fig:lrtfr_overview}
\end{figure}

As illustrated in Fig.~\ref{fig:lrtfr_overview}, for the 3D case with inputs $x_1,x_2,x_3$, the latent representations are
\begin{equation}
  \mathbf{u}=f_{\theta_1}(x_1)\in\mathbb{R}^{r_1},\;
  \mathbf{v}=f_{\theta_2}(x_2)\in\mathbb{R}^{r_2},\;
  \mathbf{w}=f_{\theta_3}(x_3)\in\mathbb{R}^{r_3}.
\end{equation}
In this case, \eqref{eq:lrtfr_tucker} reduces to the standard Tucker multilinear form
\begin{equation}
  \hat F(x_1,x_2,x_3)
  =
  \mathcal{C}\times_1 \mathbf{u}\times_2 \mathbf{v}\times_3 \mathbf{w},
\end{equation}
where $\times_n$ denotes the mode-$n$ tensor--vector product.

\subsubsection{Relevance to EM surrogates}
From the EM perspective, the appeal of LRTFR lies in the inductive bias imposed by a low-rank multilinear structure.
Many parameterized EM responses vary smoothly with respect to individual design variables, and their high-dimensional response surfaces often exhibit redundancy due to physical constraints and correlated geometric effects.
By constructing coordinate-wise latent factors and coupling them through a low-rank core, LRTFR can provide a compact representation that improves generalization when training data are limited.

\subsubsection{Limitations in high-dimensional settings}
Although LRTFR offers a principled low-rank baseline, Tucker-type coupling is not ideal when the input dimension becomes large.
On the one hand, the parameter and memory cost of the Tucker core grows exponentially with the number of dimensions, making it difficult to maintain expressive ranks in high-dimensional design spaces\cite{cichocki2015tensor}.
On the other hand, EM responses often exhibit non-uniform interactions across variables, and a single global core couples all modes in the same bottlenecked manner, which can be restrictive for representing diverse interaction patterns efficiently\cite{khrulkov2017expressive}.

\subsection{Tensor-network coupling: TT and TR baselines}
\label{subsec:tt_tr}

To obtain more scalable low-rank coupling structures, tensor-network formats replace a single global high-order core by structured contractions of matrix-like cores.
In this work, we consider tensor-train (TT)\cite{oseledets2010tt} and tensor-ring (TR) \cite{oseledets2011tensor,zhao2019learning,zhao2016tensor} as representative tensor-network couplings and implement them as additional low-rank baselines\cite{novikov2015tensorizing,stoudenmire2016supervised,cohen2016convolutional}.

\begin{figure}[!t]
  \centering
  \includegraphics[width=0.95\linewidth]{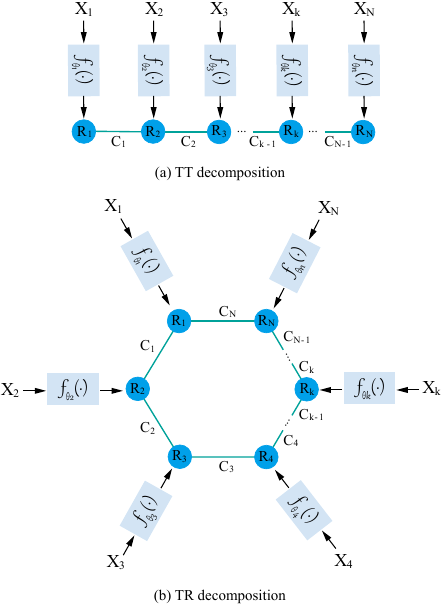}
  \caption{Tensor-network low-rank couplings used as baselines. In both cases, each input $x_i$ is mapped to $R_i(x_i)$ and reshaped into a matrix core $C_i(x_i)$.
  (a) TT coupling uses a chain-structured contraction. (b) TR coupling uses a cyclic contraction on a ring.}
  \label{fig:tttr_overview}
\end{figure}

For consistency with Fig.~\ref{fig:tttr_overview}, we denote by $R_i(x_i)$ the output of the coordinate-wise mapping before reshaping, and by $C_i(x_i)$ the reshaped matrix-valued core used in the contraction.
Importantly, $C_i(x_i)$ is input-dependent: it is generated from $x_i$ through the coordinate-wise mapping and reshaping operation, rather than being an independent learnable tensor.

\subsubsection{Tensor-train (TT)}
In TT coupling, each coordinate produces a matrix-valued core, and the output is obtained by sequential contraction along the chain.
Conceptually, TT can be viewed as distributing the global coupling into a series of low-order cores, leading to improved scaling behavior in higher dimensions.

\subsubsection{Tensor-ring (TR)}
TR coupling forms a cycle by connecting the two ends of the TT chain.
Each coordinate again produces a matrix-valued core, and the cyclic contraction aggregates information from all coordinates in a symmetric ring structure, followed by a scalar reduction.
This ring structure removes boundary effects and can provide additional flexibility compared with a pure chain.

Overall, these baselines highlight a trade-off between scalability and interaction flexibility.
This motivates a surrogate that remains low-rank and scalable, while modeling heterogeneous cross-parameter couplings more directly.

\section{PLRNet: A Pairwise Low-Rank Tensor Network for EM Surrogates}
\label{sec:method}

\subsection{From Low-Rank Baselines to PLRNet}
\label{subsec:baseline_takeaway}

Section~\ref{sec:lowrank_baselines} suggests that low-rank coupling can improve parameter efficiency for EM surrogates.
However, Tucker-type coupling can become a global bottleneck in high dimensions.
In contrast, TT/TR couplings scale more favorably but impose a fixed contraction structure.
Motivated by this gap, we propose PLRNet.
PLRNet uses compact coordinate-wise embeddings and models cross-parameter dependence through learnable pairwise interaction factors, enabling flexible interaction modeling with favorable scaling.

We aim to approximate a scalar-valued response function
\begin{equation}
  y = F(x_1,\dots,x_N), \qquad (x_1,\dots,x_N)\in\mathbb{R}^N,
\end{equation}
evaluated by an expensive electromagnetic (EM) simulator.
From the simulator, we obtain a dataset
\begin{equation}
  \mathcal{D} = \big\{(\boldsymbol{x}^{(s)}, y^{(s)})\big\}_{s=1}^m,
\end{equation}
where $\boldsymbol{x}^{(s)} = (x_1^{(s)}, \dots, x_N^{(s)})^\top$ denotes the $s$-th input sample,
$y^{(s)}$ is the corresponding EM response, and $m$ is the total number of samples.

\subsection{One-dimensional embedding networks}

\begin{figure*}[!t]
  \centering
  \includegraphics[width=\textwidth]{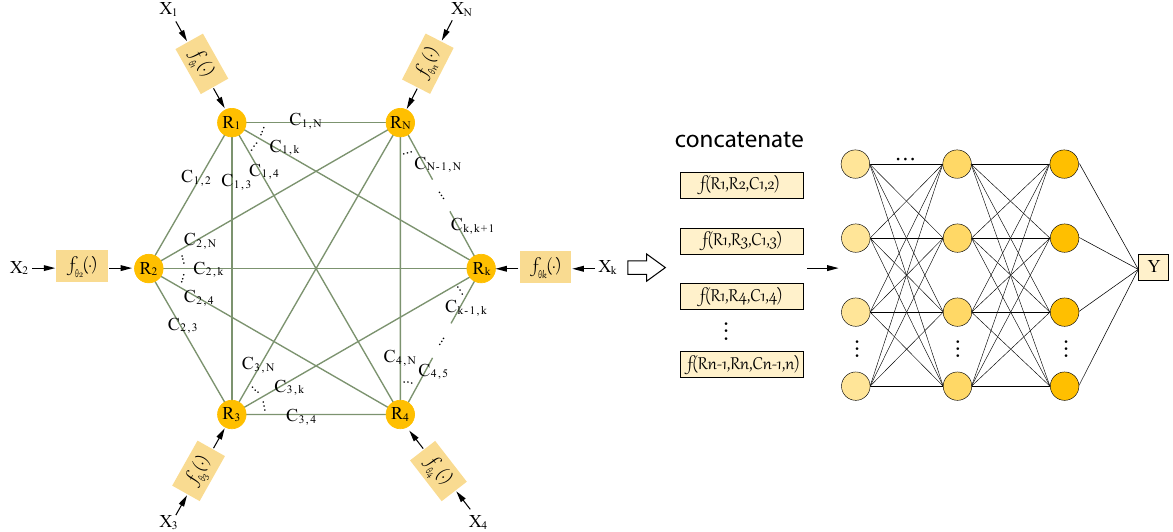}
  \caption{Overview of PLRNet. Each scalar input $x_i$ is mapped by a one-dimensional embedding network $f_{\theta_i}(\cdot)$ to a latent vector $R_i$. For each unordered pair $(i,j)$, a learnable interaction core $C_{i,j}$ produces a pairwise feature $z_{i,j}=f(R_i,R_j,C_{i,j})$. All pairwise features are concatenated and fed into a global prediction network to output the target response $y$.}
  \label{fig:plrnet_overview}
\end{figure*}

PLRNet first constructs a one-dimensional embedding network for each input component.
For the $i$-th dimension, we define a sine-activated multilayer perceptron (MLP)\cite{oseledets2013constructive}
\begin{equation}
  f_{\theta_i}:\mathbb{R}\rightarrow\mathbb{R}^{r_i},
\end{equation}
and apply it to the scalar input $x_i^{(s)}$ to obtain a latent embedding
\begin{equation}
  R_i^{(s)} = f_{\theta_i}(x_i^{(s)}) \in \mathbb{R}^{r_i}.
\end{equation}
Here, $R_i^{(s)}$ corresponds to the node $R_i$ in Fig.~\ref{fig:plrnet_overview},
representing the sample-wise latent representation of the $i$-th input dimension.

For notational completeness, stacking all samples yields a latent matrix
\begin{equation}
  \mathbf{R}_i =
  \begin{bmatrix}
    (R_i^{(1)})^\top\\
    \vdots\\
    (R_i^{(m)})^\top
  \end{bmatrix}
  \in \mathbb{R}^{m\times r_i},
\end{equation}
which is used during training but does not appear explicitly in the model architecture.
Each $f_{\theta_i}$ is implemented as a SIREN-style sine-activated MLP,
consisting of several sine layers followed by a final linear projection\cite{sitzmann2020implicit,tancik2020fourier,rahaman2019spectral}.

\subsection{Pairwise low-rank interaction cores}

To capture cross-parameter couplings, PLRNet introduces a learnable low-rank interaction
core for each unordered pair of input dimensions.
For $1 \le i < j \le N$, we define
\begin{equation}
  C_{i,j} \in \mathbb{R}^{r_i \times r_j},
\end{equation}
which corresponds to the edge labeled $C_{i,j}$ in Fig.~\ref{fig:plrnet_overview}.
Given the latent embeddings $R_i^{(s)}$ and $R_j^{(s)}$ for sample $s$,
the corresponding pairwise feature is computed via the bilinear form
\begin{equation}
  z_{i,j}^{(s)} = (R_i^{(s)})^\top C_{i,j} R_j^{(s)}.
\end{equation}
This operation explicitly models second-order interactions between different input dimensions
through a low-rank parameterization.

\subsection{Concatenation and global prediction network}

For each sample $s$, all pairwise features are collected and concatenated into a single vector
\begin{equation}
  \mathbf{z}^{(s)} =
  \big[z_{1,2}^{(s)},\, z_{1,3}^{(s)},\, \dots,\, z_{N-1,N}^{(s)}\big]^\top
  \in \mathbb{R}^{P},
  \;
  P = \frac{N(N-1)}{2}.
\end{equation}
A global prediction network
\begin{equation}
  g:\mathbb{R}^{P} \rightarrow \mathbb{R}
\end{equation}
maps the concatenated pairwise features to the final output
\begin{equation}
  \hat y^{(s)} = g\big(\mathbf{z}^{(s)};\psi\big),
\end{equation}
where $\psi$ denotes the parameters of the global predictor.
The network $g$ is implemented as a shallow SIREN-style MLP.

Overall, the proposed PLRNet model can be summarized as
\begin{equation}
  \hat F(\boldsymbol{x})
  =
  g\!\left(
  \left[\, (R_i)^\top C_{i,j} R_j \,\right]_{1 \le i < j \le N};\, \psi
  \right),\;
  R_i = f_{\theta_i}(x_i).
\end{equation}

\subsection{Training and hyper-parameter search}

PLRNet is trained by minimizing the mean-squared error
\begin{equation}
  \mathcal{L}(\Theta)
  =
  \frac{1}{m}\sum_{s=1}^{m}
  \big(y^{(s)} - \hat y^{(s)}\big)^2,
\end{equation}
where $\Theta = \{\theta_i, C_{i,j}, \psi\}$ collects all learnable parameters.

For each dataset, the samples were randomly shuffled and split into training and test sets with a ratio of 7:3. Input and output variables were standardized using the statistics of the training set. The model was trained by minimizing the mean-squared error with AdamW\cite{loshchilov2017decoupled}. To reduce overfitting, an early-stopping strategy was adopted, and the checkpoint achieving the best performance during training was retained for final evaluation.

\section{Experiments and Results}
\label{sec:experiments}

We evaluate the proposed PLRNet on electromagnetic (EM) surrogate benchmarks of increasing realism:
(i) a canonical 2D bistatic scattering task,
(ii) a parameterized microwave transmission-line return-loss task,
and (iii) a challenging 3D full-wave helix slow-wave-structure (SWS) dataset.
This progression is designed to validate two questions: (1) whether structured low-rank function representations provide an effective inductive bias on controlled EM problems, and (2) whether the resulting surrogates remain accurate and robust as the input dimension and coupling complexity increase. All models are trained with the AdamW optimizer.
We use Bayesian optimization (Optuna) to select hyper-parameters, including the embedding dimensions $r_i$, network widths and depths, SIREN frequency parameters, and the learning rate.
Each model is then retrained using the selected hyper-parameters with early stopping based on validation loss~\cite{akiba2019optuna}.
We report test mean relative error (MRE) and test maximum relative error (MaxRE) to reflect both average and worst-case performance.
We also report trainable parameters and training epochs to characterize parameter efficiency and optimization behavior.

\subsection{Bistatic RCS Prediction of an Elliptic Cylinder}
\label{subsec:rcs}

\begin{figure}[!t]
  \centering
  \includegraphics[scale=0.8]{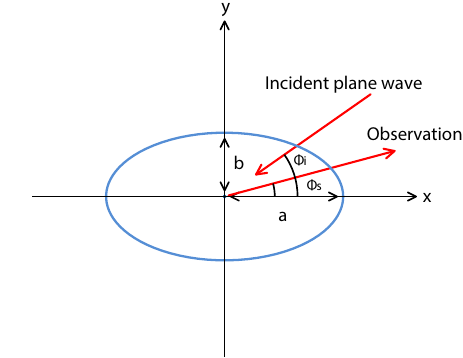}
  \caption{Elliptic-cylinder bistatic RCS benchmark. Inputs include geometry $(a,b)$ and excitation/observation settings $(f,\phi_i,\phi_s)$; the surrogate predicts $\sigma_{\mathrm{dB}}(\phi_s)$.}
  \label{fig:rcs_model}
\end{figure}

We first consider a two-dimensional scattering benchmark in which a plane wave illuminates an elliptic cylinder and the bistatic radar cross section (RCS) is evaluated at different observation angles (Fig.~\ref{fig:rcs_model})\cite{xiao2020efficient,sebak2002scattering}.
The ellipse is parameterized by its semi-major axis $a$ and semi-minor axis $b$.
For each sample, the operating frequency $f$, incidence angle $\phi_i$, and observation angle $\phi_s$ are specified.
The surrogate target is the RCS in dB, denoted as $\sigma_{\mathrm{dB}}(\phi_s)$.

The supervised mapping is five-dimensional:
\begin{equation}
  (a,\, b,\, f,\, \phi_i,\, \phi_s)\ \mapsto\ \sigma_{\mathrm{dB}}(\phi_s).
\end{equation}
The sampling ranges and dataset size are summarized in Table~\ref{tab:rcs_data}.
A total of $87{,}360$ samples are generated, covering variations in geometry, frequency, and angular configurations.

\begin{table}[!t]
\centering
\caption{Sampling ranges for the elliptic-cylinder RCS dataset.}
\label{tab:rcs_data}
\begin{tabular}{cc}
\toprule
Variable & Range \\
\midrule
$a$ (m) & [2.00, 2.34] \\
$b$ (m) & [1.00, 1.70] \\
$f$ (GHz) & [0.30, 0.52] \\
$\phi_i$ (deg) & [0.00, 60.00] \\
$\phi_s$ (deg) & [0.00, 31.00] \\
\midrule
Total samples & 87{,}360 \\
\bottomrule
\end{tabular}
\end{table}

We compare PLRNet against a generic MLP and representative low-rank baselines (LRTFR, TT, and TR).

\begin{table}[!t]
\centering
\caption{Performance comparison on the RCS benchmark.}
\label{tab:rcs_results}
\setlength{\tabcolsep}{6pt}
\begin{tabular}{lcccc}
\toprule
Model & Params & Test MRE & Test MaxRE & Epoch \\
\midrule
PLRNet & \textbf{\num{2.01e6}} & \textbf{\num{3.78e-4}} & \textbf{\num{9.02e-3}} & 10000 \\
TR     & \num{5.31e6} & \num{5.38e-4} & \num{3.36e-2} & 8000 \\
MLP    & \num{6.79e6} & \num{6.47e-4} & \num{1.36e-2} & 10000 \\
TT     & \num{2.39e6} & \num{6.59e-4} & \num{2.15e-2} & 8000 \\
LRTFR  & \num{7.97e6} & \num{1.26e-3} & \num{5.98e-2} & 4287 \\
\bottomrule
\end{tabular}
\end{table}

\begin{figure}[!t]
  \centering
  \includegraphics[scale=0.6]{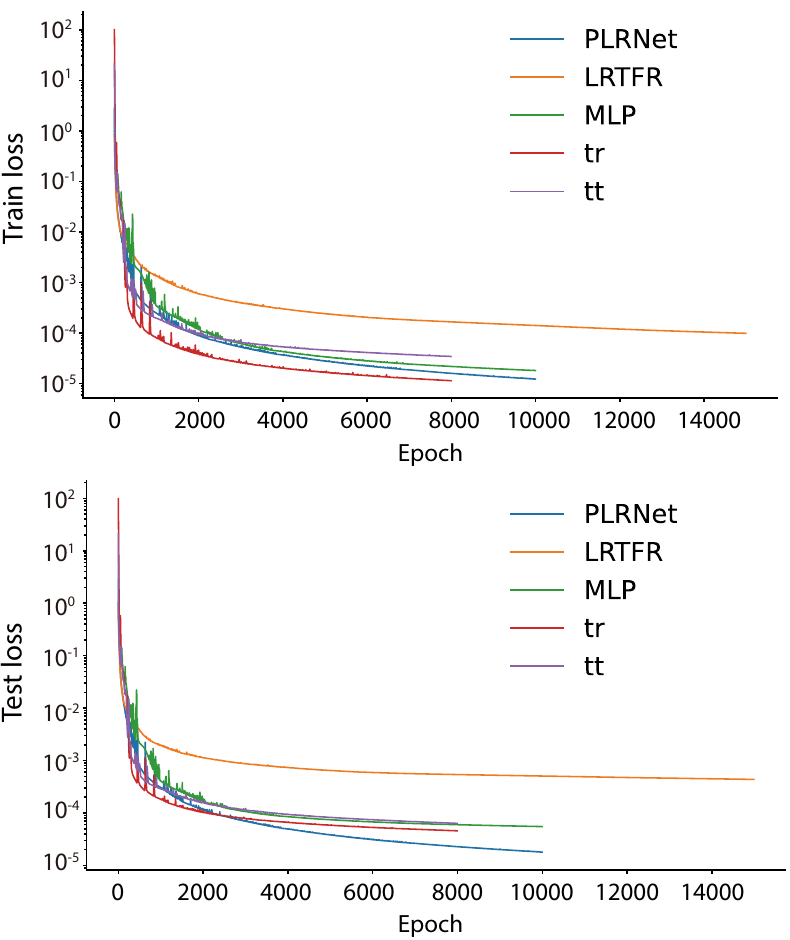}
  \caption{Training and test loss curves for the elliptic-cylinder RCS benchmark.}
  \label{fig:rcs_loss}
\end{figure}

The training dynamics are first shown in Fig.~\ref{fig:rcs_loss}.
Overall, the low-rank structured models exhibit stable optimization behavior on this benchmark, which supports the use of low-rank decomposition as a practical surrogate-modeling methodology for scattering problems.
Among them, PLRNet converges smoothly and reaches a lower final loss than the competing methods, suggesting that its pairwise interaction structure provides an optimization advantage in addition to parameter compression.

\begin{figure}[!t]
  \centering
  \includegraphics[scale=0.6]{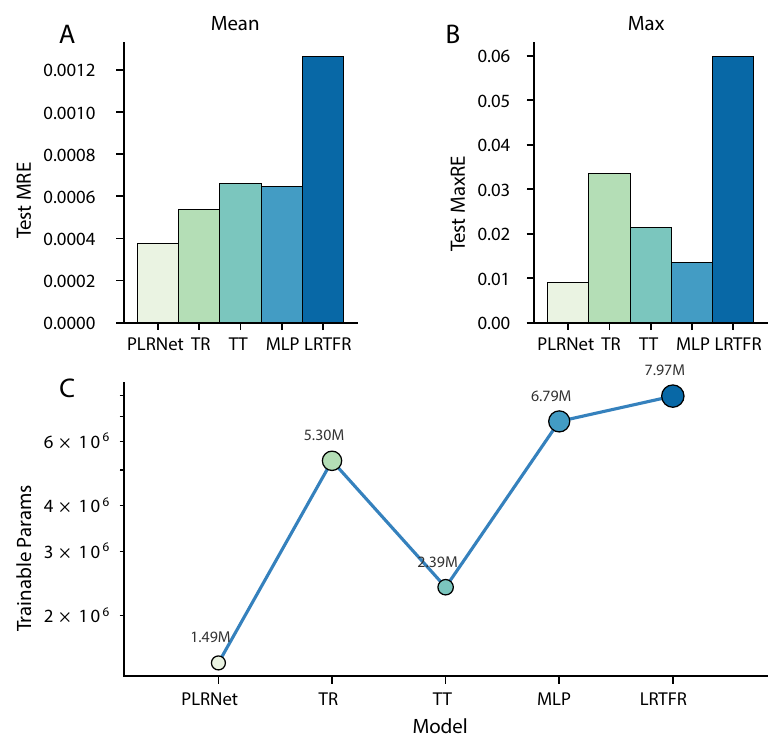}
  \caption{RCS benchmark results across different models. (A) Mean relative error (MRE). (B) Maximum relative error (MaxRE). (C) Trainable parameters, where marker size reflects model complexity.}
  \label{fig:rcs_results}
\end{figure}

The quantitative comparison is summarized in Fig.~\ref{fig:rcs_results} and Table~\ref{tab:rcs_results}.
Several observations can be drawn.
First, this benchmark already illustrates the value of low-rank structured surrogate modeling for EM scattering problems.
Compared with the dense MLP baseline, the low-rank models generally achieve comparable or better prediction accuracy with substantially fewer parameters, indicating that the underlying RCS mapping contains exploitable low-dimensional structure.
For example, TR reduces the average error compared with MLP (test MRE \num{5.38e-4} vs.\ \num{6.47e-4}), while TT attains a very similar MRE to MLP using far fewer parameters (\num{2.39e6} vs.\ \num{6.79e6}).
These results support the broader methodological claim that low-rank decomposition provides a useful inductive bias for structured EM response mappings.

Second, the representative low-rank baselines exhibit different trade-offs.
TT is particularly parameter-efficient, suggesting that tensor-network coupling can remove redundancy effectively even when the accuracy gain is modest.
TR achieves a lower mean error than MLP, but its worst-case error is noticeably larger (test MaxRE \num{3.36e-2}), which indicates that a fixed ring-structured contraction may not always align with the most critical interaction patterns in the scattering response.
LRTFR performs worst in both average and worst-case metrics, suggesting that a single Tucker-style global core is less suitable for this task, where the interactions across geometry, excitation, and observation variables are heterogeneous.

Third, within this low-rank modeling family, PLRNet achieves the most favorable overall balance.
It attains the lowest test MRE (\num{3.78e-4}) and the lowest test MaxRE (\num{9.02e-3}), while also using the fewest trainable parameters (\num{2.01e6}) among all compared models.
This indicates that, beyond the general benefit of low-rank decomposition, the pairwise interaction structure of PLRNet is better matched to the dominant coupling patterns in this RCS task.
In particular, the improvement in MaxRE suggests stronger robustness across angular configurations, which is important because scattering responses can vary sharply with the observation angle $\phi_s$.

Taken together, this benchmark suggests that low-rank decomposition is already a useful modeling principle for moderate-dimensional scattering problems, while PLRNet further improves the accuracy--efficiency trade-off through an EM-oriented pairwise interaction design.

\subsection{Return-Loss Prediction of a Loaded Microstrip Line}
\label{subsec:txline}

\begin{figure}[!t]
  \centering
  \includegraphics[scale=0.8]{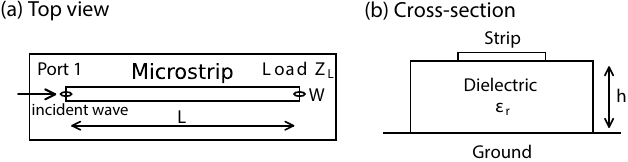}
  \caption{Loaded microstrip-line benchmark: (a) top view showing Port~1, line length $L$, the resistive load $Z_L$, and strip width $W$; (b) cross-section showing substrate permittivity $\varepsilon_r$, ground plane, and substrate thickness $h$. The surrogate predicts the input return loss $RL_{\mathrm{dB}}$.}
  \label{fig:txline_model}
\end{figure}

We next evaluate the surrogates on a microwave transmission-line benchmark: a microstrip line terminated by a resistive load (Fig.~\ref{fig:txline_model})\cite{hammerstad1980accurate}.
The prediction target is the input return loss at the excitation port, a standard metric for impedance matching in microwave circuits\cite{zhang2018multivalued}.
The structure consists of a strip conductor of width $W$ printed on a dielectric substrate of thickness $h$ and relative permittivity $\varepsilon_r$, backed by a ground plane.
The line length is $L$, the termination is $Z_L$, and the operating frequency is $f$.

The supervised mapping is
\begin{equation}
  (W,\, h,\, \varepsilon_r,\, L,\, f,\, Z_L)\ \mapsto\ RL_{\mathrm{dB}}.
\end{equation}
The sampling ranges are summarized in Table~\ref{tab:txline_data}, with $6{,}000$ samples in total.
Although the structure admits transmission-line interpretations, the return loss still involves nonlinear coupled effects among geometry, material, and frequency-dependent quantities, for example through effective permittivity and characteristic impedance, making it a meaningful test for interaction modeling\cite{kirschning2003accurate}.

\begin{table}[!t]
\centering
\caption{Sampling ranges for the loaded microstrip-line dataset.}
\label{tab:txline_data}
\begin{tabular}{cc}
\toprule
Variable & Range \\
\midrule
$W$ (mm) & [0.2, 0.5] \\
$h$ (mm) & [0.2, 0.5] \\
$\varepsilon_r$ & [2.2, 3.0] \\
$L$ (mm) & [1.0, 5.0] \\
$f$ (GHz) & [5.0, 7.0] \\
$Z_L$ ($\Omega$) & [20.0, 22.0] \\
\midrule
Total samples & 6{,}000 \\
\bottomrule
\end{tabular}
\end{table}

Table~\ref{tab:txline_results} summarizes the performance comparison.

\begin{table}[!t]
\centering
\caption{Performance comparison on the microstrip return-loss benchmark.}
\label{tab:txline_results}
\setlength{\tabcolsep}{6pt}
\begin{tabular}{lcccc}
\toprule
Model & Params & Test MRE & Test MaxRE & Epoch \\
\midrule
PLRNet & \num{2.97e6} & \textbf{\num{9.90e-5}} & \textbf{\num{7.74e-4}} & 10000 \\
TR     & \num{2.03e6} & \num{5.30e-4} & \num{5.17e-3} & 3800 \\
MLP    & \num{1.54e7} & \num{6.65e-4} & \num{4.44e-3} & 5440 \\
TT     & \textbf{\num{1.31e6}} & \num{1.73e-3} & \num{2.49e-2} & 4000 \\
LRTFR  & \num{2.30e7} & \num{1.82e-3} & \num{2.51e-2} & 7235 \\
\bottomrule
\end{tabular}
\end{table}

\begin{figure}[!t]
  \centering
  \includegraphics[scale=0.65]{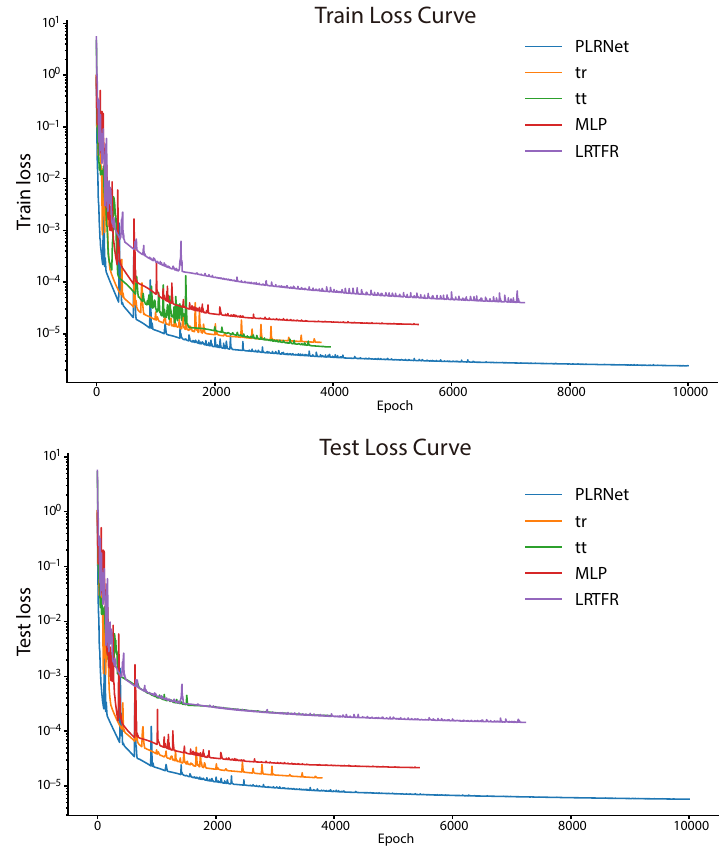}
  \caption{Training and test loss curves for the loaded microstrip-line benchmark.}
  \label{fig:txline_loss}
\end{figure}

The training dynamics are first illustrated in Fig.~\ref{fig:txline_loss}.
Overall, the low-rank structured surrogates optimize stably on this benchmark, again supporting low-rank decomposition as a viable modeling methodology for EM responses with structured parameter couplings.
PLRNet shows the smoothest convergence and the lowest final test loss, indicating that its interaction-aware pairwise design not only improves representational efficiency but also facilitates optimization.

\begin{figure}[!t]
  \centering
  \includegraphics[scale=0.6]{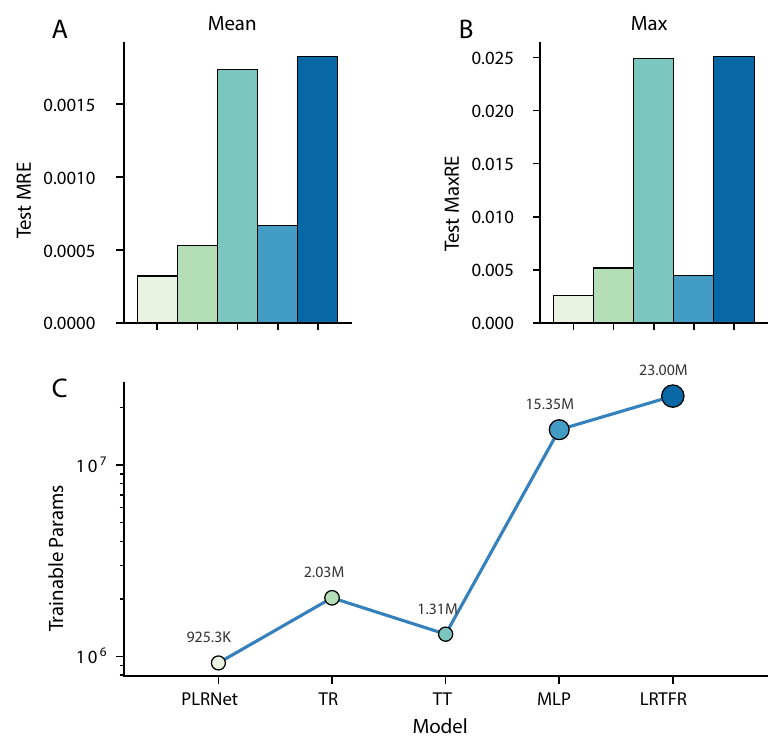}
  \caption{Microstrip benchmark results across different models. (A) Mean relative error (MRE). (B) Maximum relative error (MaxRE). (C) Trainable parameters, with marker size reflecting model complexity.}
  \label{fig:chuanshuxian_results}
\end{figure}

The quantitative comparison is then presented in Fig.~\ref{fig:chuanshuxian_results} and Table~\ref{tab:txline_results}.
This benchmark further supports low-rank modeling as an effective surrogate-design principle for EM responses with structured parameter couplings.
Compared with the dense MLP, the low-rank models achieve competitive or better performance with markedly smaller parameter budgets in most cases, indicating that the return-loss mapping is not only nonlinear but also structurally compressible.
For instance, TR improves the average accuracy over MLP (test MRE \num{5.30e-4} vs.\ \num{6.65e-4}) using far fewer parameters, and TT is the most compact model overall, with only \num{1.31e6} trainable parameters.
These observations again support the broader methodological claim that low-rank decomposition is a suitable inductive bias for this class of EM surrogate problems.

At the same time, the representative low-rank baselines reveal different inductive biases.
TR is competitive in terms of average error, but its test MaxRE is larger than that of PLRNet and MLP, indicating that certain parameter combinations remain difficult under a fixed ring-structured coupling.
TT offers the strongest compression, but it shows a clear degradation in both MRE and MaxRE, suggesting that a fixed sequential contraction order can be restrictive when the dominant couplings in the return-loss mapping are not well aligned with a chain structure.
LRTFR again yields relatively large errors while also being the largest model on this task, reinforcing that Tucker-style global coupling is not the most parameter-efficient choice for this type of multi-parameter EM response.

Within this broader low-rank family, PLRNet achieves the best overall performance.
It obtains the lowest test MRE (\num{9.90e-5}) and the lowest test MaxRE (\num{7.74e-4}), indicating excellent average accuracy and strong worst-case robustness.
Compared with MLP, PLRNet reduces the MRE by roughly an order of magnitude while using far fewer parameters (\num{2.97e6} vs.\ \num{1.54e7}).
These results suggest that the benefit comes not only from using a compact low-rank representation, but also from adopting a pairwise interaction structure that better matches the coupled dependence among geometry, material, load, and frequency variables in the microstrip problem.

Overall, this benchmark again demonstrates the usefulness of low-rank structured surrogates over dense baselines, while showing that an EM-tailored pairwise low-rank design can further improve the accuracy--efficiency trade-off.

\subsection{Surrogate Prediction for Helix Slow-Wave Structure Responses}
\label{subsec:helix}

\begin{figure}[!t]
  \centering
  \begin{minipage}{.24\textwidth}
    \centering
    \includegraphics[scale=0.222]{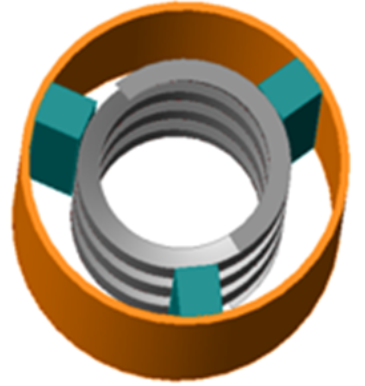}
  \end{minipage}
  \begin{minipage}{0.24\textwidth}
    \centering
    \includegraphics[scale=0.3]{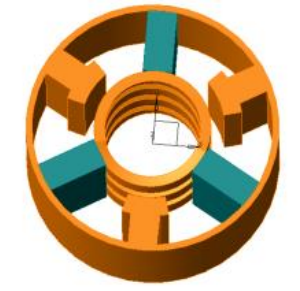}
  \end{minipage}
  \caption{3D model of a helix slow-wave structure (SWS) in a traveling-wave tube (TWT).}
  \label{fig:sws_geometry}
\end{figure}

We further evaluate PLRNet on a realistic and challenging 3D full-wave benchmark, a helix slow-wave structure (SWS),
which is the core interaction circuit in traveling-wave tubes (TWTs), as illustrated in Fig.~\ref{fig:sws_geometry}\cite{zhao2023efficient,zhao2025efficient,paoloni2021millimeter}.
Unlike the previous two comparatively controlled benchmarks,
this dataset is obtained from high-fidelity full-wave EM simulations, which capture strong nonlinear and heterogeneous cross-parameter couplings\cite{duan2007impact}.
We consider three prediction tasks from the simulated responses, including phase velocity, interaction impedance, and attenuation.

To study scalability, we construct three datasets with increasing input dimensionality (4D, 7D, and 10D) and different sample sizes.
The parameter ranges are summarized in Table~\ref{tab:sws_ranges}.
The surrogate outputs are sampled over frequency, and the resulting mapping is highly nonlinear with strong cross-parameter interactions.
This setting directly reflects practical full-wave surrogate modeling where sample budgets are limited for low-dimensional subsets and the interaction structure becomes more complex as the dimension increases.

\begin{table}[!t]
\centering
\caption{Value ranges of variables in the 4D, 7D, and 10D helix SWS datasets.}
\label{tab:sws_ranges}
\setlength{\tabcolsep}{4pt}
\begin{tabular}{cccc}
\toprule
Variable & 4D & 7D & 10D \\
\midrule
Pitch(mm)     & [0.7, 0.9]     & [0.81, 0.99]   & [0.531, 0.649] \\
helixRa(mm)   & [1.11, 1.19]   & [0.72, 0.88]  & [0.594, 0.726] \\
JW(mm)        & [0.47, 0.55]   & [0.36, 0.44]   & [0.36, 0.44] \\
Frequency(GHz) & [4, 12]        & [11, 13]   & [6, 18] \\
shellRc(mm)   & --             & [1.351, 1.65] & [1.485, 1.815] \\
helixHd(mm)   & --             & [0.18, 0.22]   & [0.126, 0.154] \\
helixHw(mm)   & --             & [0.45, 0.55]   & [0.324, 0.396] \\
YRL(mm)       & --             & --             & [0.891, 1.089] \\
YRs(mm)       & --             & --             & [1.116, 1.364] \\
YVn(mm)       & --             & --             & [0.36, 0.44] \\
\midrule
Total samples & 750 & 2345 & 7423 \\
\bottomrule
\end{tabular}
\end{table}

We report the parameter count in Table~\ref{tab:sws_params} to highlight model scalability under increasing input dimensions and task settings.
``--'' indicates that LRTFR\cite{luo2023low} is not evaluated on 10D due to the exponential growth of the Tucker core with dimension.

\begin{table*}[!t]
\centering
\caption{Number of trainable parameters under different feature dimensions on the helix SWS tasks.}
\label{tab:sws_params}
\begin{tabular}{lccccccccc}
\toprule
\multirow{2}{*}{Model}
& \multicolumn{3}{c}{Phase velocity}
& \multicolumn{3}{c}{Interaction impedance}
& \multicolumn{3}{c}{Attenuation} \\
\cmidrule(lr){2-4} \cmidrule(lr){5-7} \cmidrule(lr){8-10}
& 4D & 7D & 10D
& 4D & 7D & 10D
& 4D & 7D & 10D \\
\midrule
PLRNet & 1.53M & \textbf{0.80M} & \textbf{2.04M} & \textbf{0.41M} & 1.07M & \textbf{1.65M} & 0.56M & 1.62M & \textbf{1.19M} \\
TR     & \textbf{0.87M} & 3.74M & 6.10M & 0.70M & \textbf{0.07M} & 3.18M & \textbf{0.05M} & 2.83M & 2.95M \\
TT     & 1.31M & 4.15M & 8.11M & 3.02M & 0.43M & 9.68M & 1.33M & \textbf{1.33M} & 2.02M \\
MLP    & 11.94M & 6.05M & 18.97M & 28.67M & 44.77M & 40.49M & 11.37M & 45.95M & 65.41M \\
LRTFR  & 3.58M & 424.78M & --    & 2.73M & 4.78M & --    & 0.65M & 5.54M & -- \\
\bottomrule
\end{tabular}
\end{table*}

Table~\ref{tab:sws_params} shows a clear difference in scaling behavior.
MLP becomes very large as the task and dimension change, reaching 65.41M parameters on 10D attenuation.
LRTFR also grows rapidly with dimension and is not available on 10D.
In comparison, PLRNet remains within about 0.41M to 2.04M parameters across all tasks and dimensions, which makes it attractive for multi-task full-wave surrogate learning.

\begin{figure*}[!t]
  \centering
  \includegraphics[scale=0.85]{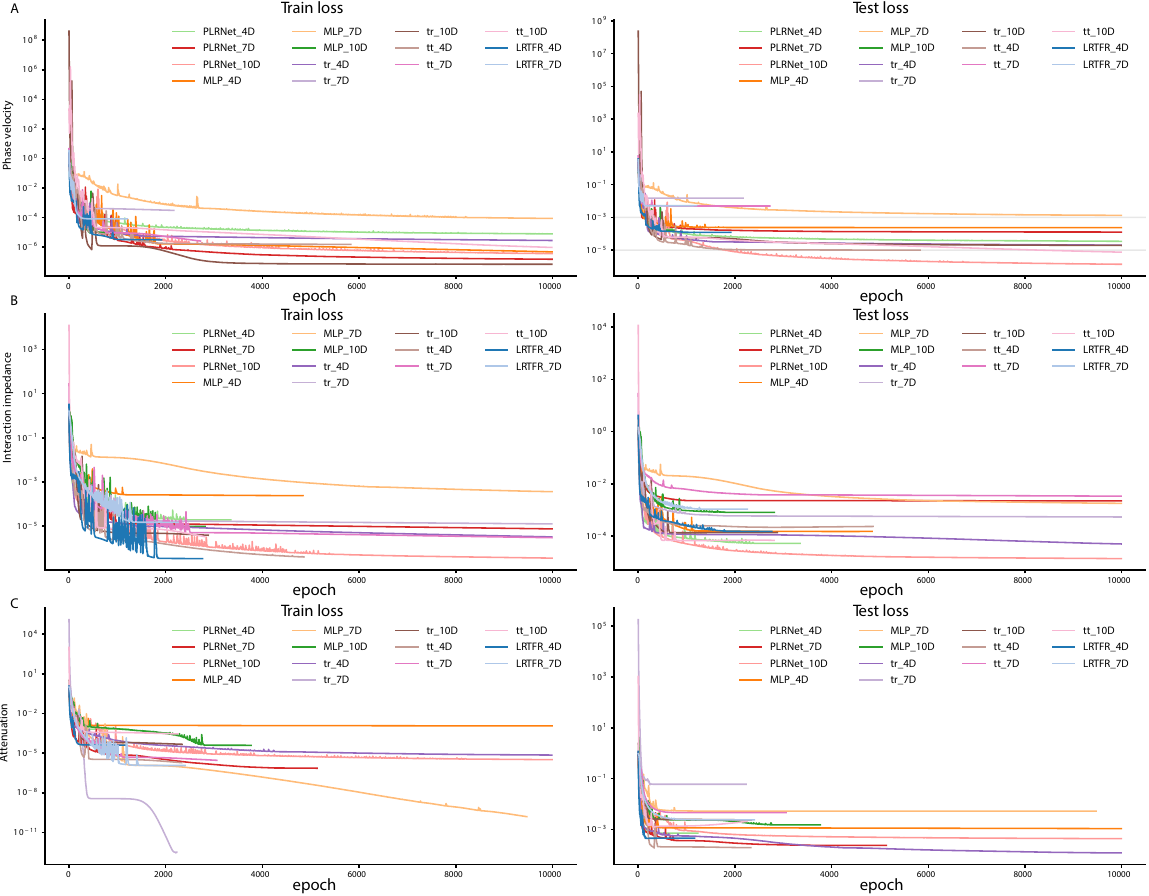}
  \caption{Training and test loss curves of different models on the helix SWS datasets (4D, 7D, and 10D) for three prediction tasks: (A) phase velocity, (B) interaction impedance, and (C) attenuation. Left column shows training loss, and right column shows test loss.}
  \label{fig:sws_loss}
\end{figure*}
\begin{figure*}[!t]
  \centering
  \includegraphics[scale=0.85]{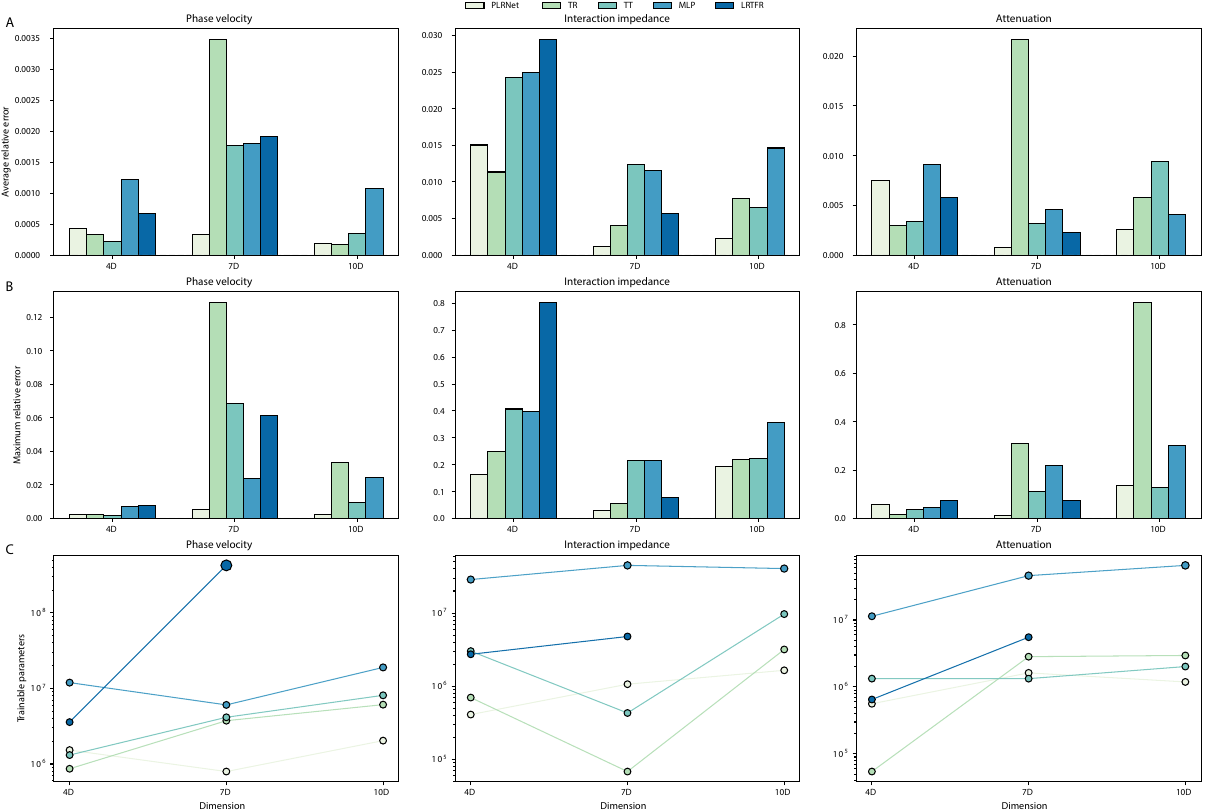}
  \caption{Performance comparison across different models on the helix SWS datasets (4D, 7D, and 10D). Columns correspond to three prediction tasks: phase velocity, interaction impedance, and attenuation. Rows correspond to (A) mean relative error (MRE), (B) maximum relative error (MaxRE), and (C) trainable parameters.}
  \label{fig:sws_results}
\end{figure*}
The learning curves in Fig.~\ref{fig:sws_loss}(A)--(C) indicate stable optimization for PLRNet on all three tasks.
Its training loss decreases smoothly and the corresponding test loss converges to a low level without obvious oscillation.
Several baselines plateau earlier with higher final test loss, and this tendency becomes more visible on the higher-dimensional settings.

The quantitative comparison is summarized in Fig.~\ref{fig:sws_results}(A)--(C), which report the test MRE, test MaxRE, and trainable parameters, respectively. Overall, the results support the effectiveness of low-rank modeling as a general methodology for high-dimensional electromagnetic surrogate learning. Compared with the dense MLP baseline, the low-rank models, including TT, TR, Tucker-based LRTFR, and PLRNet, usually achieve competitive or better accuracy with substantially fewer trainable parameters, indicating that exploiting low-dimensional structure in the parameter space is beneficial for these full-wave prediction tasks.

On the phase velocity task, the low-rank models consistently perform strongly across all feature dimensions. TT achieves the lowest MRE on 4D (0.000226), PLRNet performs best on 7D with MRE 0.000338 and MaxRE 0.005077, and TR attains the lowest MRE on 10D (0.000172), slightly better than PLRNet (0.000182). These results suggest that for relatively smooth targets such as phase velocity, generic low-rank decompositions already provide highly competitive approximations. At the same time, PLRNet shows improved robustness in the more challenging settings, especially on 10D, where it achieves a much smaller MaxRE (0.002183) than TR (0.033182) and MLP (0.024608). LRTFR also benefits from the low-rank modeling principle, but its Tucker-style parameterization becomes much less scalable at higher dimensions, as reflected by the exceptionally large parameter count on 7D phase velocity.

On the interaction impedance task, the advantage of structured low-rank modeling becomes more evident. TR gives the lowest MRE on 4D (0.011344), but PLRNet already achieves the lowest MaxRE (0.162178) there. More importantly, on 7D and 10D, PLRNet clearly outperforms the other models in both mean and worst-case errors, with MRE/MaxRE of 0.001102/0.028549 on 7D and 0.002321/0.192503 on 10D. Since interaction impedance is more sensitive to coupled variations among geometry and frequency parameters, these results indicate that explicitly structured low-rank representations are particularly suitable for capturing such dependencies. TT, TR, and LRTFR all demonstrate the usefulness of low-rank decomposition compared with dense baselines, while PLRNet further improves the results by incorporating a problem-aware interaction structure.

A similar trend is observed on the attenuation task. In 4D, TR performs best, achieving the lowest MRE (0.002934) and MaxRE (0.016337), which again shows that standard low-rank decomposition is already highly effective in lower-dimensional settings. However, as the dimensionality increases, PLRNet becomes the most competitive model. On 7D, it achieves the best MRE (0.000791) and MaxRE (0.012349), and on 10D it attains the lowest MRE (0.002527), while TT obtains a slightly smaller MaxRE (0.128378 versus 0.136753 for PLRNet). Nevertheless, TT has a much larger mean error (0.009388), so PLRNet provides the best overall trade-off between average accuracy and robustness. This again supports the view that low-rank decomposition is an effective modeling principle, while the PLRNet formulation better aligns with the cross-parameter coupling structure of the electromagnetic problem.

From the model-efficiency perspective, the results further highlight the practical value of low-rank parameterization. MLP often requires tens of millions of trainable parameters, whereas TT, TR, and PLRNet usually operate with a much smaller parameter budget. Tucker-based LRTFR is also more compact than dense baselines in some lower-dimensional cases, but its parameter count grows rapidly as the dimensionality increases, which limits its scalability. Although PLRNet does not always use the fewest parameters in every setting, it achieves the most consistent balance among compactness, average accuracy, and worst-case error across the three tasks. In contrast, the generic low-rank baselines can be very competitive on specific tasks or dimensions, but their performance is less stable as the problem complexity increases.

Overall, the helix SWS benchmark suggests that low-rank decomposition is a promising methodology for high-dimensional electromagnetic surrogate modeling. Standard tensor low-rank models such as TT, TR, and Tucker-based LRTFR already offer clear benefits over dense baselines, especially in parameter efficiency and, in several cases, predictive accuracy. Building on this methodology, PLRNet further improves the performance by introducing a low-rank structure tailored to electromagnetic parameter interactions, leading to a more favorable and more consistent accuracy--robustness trade-off across the three prediction tasks.

\section{Conclusion}
\label{sec:conclusion}

This work advocates low-rank tensor function representations as a principled paradigm for electromagnetic (EM) surrogate modeling, aiming to achieve both accuracy and scalability in high-dimensional parameter spaces under limited simulation budgets.
Beyond introducing this perspective to EM surrogates, we provided a systematic study of representative low-rank couplings, including Tucker-style LRTFR as well as tensor-network couplings based on tensor-train (TT) and tensor-ring (TR), which serve as structured baselines and clarify practical trade-offs between expressiveness, scalability, and optimization behavior.

Building on the insights from these baselines, we proposed PLRNet, a pairwise low-rank tensor network that combines compact coordinate-wise representations with learnable pairwise interaction factors to model cross-parameter dependence in an interaction-aware yet scalable manner.
Across three benchmarks of increasing realism and difficulty, including 2D bistatic scattering, loaded microstrip return-loss prediction, and a challenging 3D full-wave helix slow-wave-structure (SWS) dataset, PLRNet outperforms dense MLPs and shows competitive or superior performance relative to representative low-rank baselines across most settings, with stable optimization in high-dimensional regimes.
These results indicate that distributing low-rank coupling over parameter pairs is an effective strategy for practical high-dimensional EM surrogate learning and can support fast evaluation in parametric studies and design-space exploration.

Several directions may further strengthen this framework.
First, the low-rank function representation viewpoint naturally extends to richer EM outputs, such as multi-frequency responses and vector-valued quantities, via multi-head prediction layers or structured output parameterizations.
Second, incorporating physics-aware priors, including passivity and causality constraints for S-parameters, symmetry constraints, and monotonicity where appropriate, may further improve reliability, especially in extrapolative regimes\cite{gustavsen2002rational}.
Finally, as the number of parameter pairs grows with dimension, PLRNet can be made more efficient in very high-dimensional settings by selecting or sparsifying interaction pairs using validation-driven screening or lightweight sensitivity analysis, which may also improve interpretability of the learned couplings.

\section*{Acknowledgement}
The last author is supported by NSFC, 12101511 and Fundamental Research Funds for the Central Universities (Grant No. JBK22YJ37).

\bibliography{References.bib}



\end{document}